\documentclass[runningheads]{llncs}
\usepackage[T1]{fontenc}
\usepackage{graphicx}
\usepackage[hyphens]{url}
\usepackage{hyperref}
\usepackage[hyphenbreaks]{breakurl}

\begin{document}
\title{\mbox{Recommender Systems for Social Good:} \mbox{The Role of Accountability and Sustainability}}
\titlerunning{Recommender Systems and the SDGs}
%
\author{Alan Said\inst{1}\orcidID{0000-0002-2929-0529}}
\authorrunning{A. Said}
%
\institute{University of Gothenburg, Sweden
\email{alansaid@acm.org}
}
\maketitle              
\begin{abstract}
 This work examines the role of recommender systems in promoting sustainability, social responsibility, and accountability, with a focus on alignment with the United Nations Sustainable Development Goals (SDGs). As recommender systems become increasingly integrated into daily interactions, they must go beyond personalization to support responsible consumption, reduce environmental impact, and foster social good. We explore strategies to mitigate the carbon footprint of recommendation models, ensure fairness, and implement accountability mechanisms. By adopting these approaches, recommender systems can contribute to sustainable and socially beneficial outcomes, aligning technological advancements with the SDGs focused on environmental sustainability and social well-being.
\end{abstract}
\section{Introduction}
Recommender systems have become integral to our digital experiences, guiding us in selecting content, products, and services across domains such as entertainment, e-commerce, and health. With their expanding influence, these systems are uniquely positioned to contribute to societal goals beyond personalization and engagement, aligning with frameworks like the United Nations Sustainable Development Goals\footnote{\url{https://sdgs.un.org/goals}}. However, this potential comes with responsibility: as these systems grow, so do concerns regarding their environmental footprint, fairness, and accountability.

This work explores how recommender systems can be aligned with SDGs focused on sustainability, social responsibility, and accountable use of technology. Specifically, it addresses the need for strategies to mitigate the carbon impact of recommender systems, enhance fairness across user groups, and establish accountability mechanisms to increase transparency and trust. These approaches support goals like SDG 10---\emph{Reduced Inequality}, SDG 12---\emph{Responsible Consumption and Production}, SDG 13---\emph{Climate Action}, and SDG 16---\emph{Peace, Justice, and Strong Institutions}, demonstrating how technological advancements in recommender systems can be steered towards social good.

\section{Recommender Systems Through the Lens of the SDGs}
Recommender systems play a significant role in shaping digital consumption, offering a pathway to support the SDGs. As these systems influence user choices and behaviors, they can either drive positive societal change or exacerbate existing challenges. Here, we explore how recommender systems can be designed to align with specific SDGs, primarily focusing on social responsibility, sustainability, and accountability.

\subsection{SDG 10: Reduced Inequality}

Fairness in recommender systems is essential to prevent the reinforcement of social biases, aligning with SDG 10’s goal of reducing inequalities. Biased algorithms in sensitive domains like criminal justice and finance can perpetuate unequal access to opportunities. For instance, automated systems in judicial sentencing \cite{polonskiAIConvictingCriminals2018} and financial scoring \cite{marshNationsLargestCredit2023} have raised concerns due to potential demographic biases, underscoring the importance of fairness in recommendation and decision-support applications with significant real-world implications. Achieving fairness in this space, however, is complex, as there is no universally agreed-upon definition. Abu Elyounes \cite{abuelyounesContextualFairnessLegal2019} notes that much of the research on fairness in AI involves justifying why one notion is ``fairer'' than others, reflecting the diversity of fairness interpretations. Li et al. \cite{liFairnessRecommendationFoundations2023} emphasize that fairness needs vary by context, with metrics like demographic parity and equalized odds presenting trade-offs that impact perceived fairness.

A study by Kecki and Said \cite{keckiUnderstandingFairnessRecommender2024} demonstrates that public perceptions of fairness are context-dependent, with preferences shifting based on the stakes involved, further highlighting the need for contextual sensitivity in fairness-aware systems. An adaptable approach to fairness is essential: high-stakes applications like healthcare may prioritize accuracy, while applications emphasizing equal access could focus on demographic parity. By adopting such adaptive, context-aware approaches, recommender systems can support SDG 10 by fostering equitable access and reducing inequalities across various domains.

\subsection{SDG 12: Responsible Consumption and Production \& SDG 13: Climate Action}

Recommender systems can support responsible consumption by guiding users towards sustainable choices. However, these systems often rely on large-scale, deep learning models with substantial environmental impacts. Recent studies show that the average carbon footprint of a single deep learning recommender system paper can exceed 3,000 kilograms of CO\textsubscript{2} equivalents---a figure comparable to long-haul flight emissions \cite{venteClicksCarbonEnvironmental2024}. 

In 2023 alone, over 17,000 papers were published on deep learning recommenders\footnote{\url{https://scholar.google.com/scholar?as\_ylo=2023\&as\_yhi=2023\&q=\%22deep+learning\%22+\%22recommender+system\%22\%7C\%22recommender+systems\%22\%7C\%22recommendation+system\%22\%7C\%22recommendation+systems\%22}}, placing the cumulative emissions of this field on par with those of smaller nations \cite{emissions}. This environmental cost prompts questions about whether high-emission models are necessary for every use case.

``Good old-fashioned AI'' models, like neighborhood-based methods, can offer competitive recommendation accuracy without deep learning’s energy demands. For some applications, the incremental accuracy of deep learning may not justify its carbon footprint. For instance, recommending digital content (e.g., music or movies) has low environmental impact even if recommendations are suboptimal, i.e., skipping to the next track or selecting a different movie causes negligible CO\textsubscript{2} emissions. However, in fashion e-commerce, poor recommendations can lead to significant environmental costs from shipping and returns, e.g., up to half of fashion e-commerce orders in Europe are reportedly returned \cite{marketingyocabeGuidaAiResi2023}. Here, accurate recommendations could reduce unnecessary shipments, directly supporting SDG 12 by encouraging responsible consumption.

To mitigate environmental impact, practitioners can explore energy-efficient algorithms, optimize resources, and consider carbon offsets. Additionally, recommender systems can align with SDG 13 by prioritizing sustainable options, such as local products or eco-friendly items, fostering environmentally conscious behaviors across different application contexts.

\subsection{SDG 16: Peace, Justice, and Strong Institutions}

Accountability and transparency in recommender systems are foundational to building trust, aligning with SDG 16's focus on strong institutions. Users benefit from knowing how and why recommendations are made, as transparency fosters trust and allows users to make informed choices. Accountability mechanisms---such as transparency, explainability, and reproducibility---enable stakeholders to monitor and audit these systems, ensuring they act fairly and justly.

One approach to promoting accountability in recommender systems is through reproducibility, which strengthens research validity and builds trust within the research and practitioner communities as well as with users \cite{belloginImprovingAccountabilityRecommender2021}. Furthermore, accountability allows for identifying and addressing biases, supporting ethical and socially responsible recommendations. This practice aligns recommender systems with the principles of justice and fair treatment, reinforcing their role as trusted digital facilitators.

\section{Conclusions}
Recommender systems research and practice offer many opportunities to support the SDGs by making choices that promote accountability, transparency, and context-aware design. Through careful consideration of these principles, the research and practitioner communities can align recommender systems with the goals of reducing inequalities (SDG 10), fostering responsible consumption and production (SDG 12), advancing climate action (SDG 13), and strengthening justice and institutions (SDG 16). However, achieving these outcomes requires a nuanced and contextual approach; there is no ``one size fits all'' solution, as different applications and environments demand tailored strategies to ensure effective and sustainable outcomes.

While this work focused on SDGs 10, 12, 13, and 16, many other SDGs can be linked to recommender systems, from promoting good health to supporting quality education and beyond. This highlights the importance of researchers and practitioners in the recommender systems field to adopt societal good as a guiding principle, continually reflecting on the broader impacts of their work on global well-being.

%
%
%
\bibliographystyle{splncs04}
\bibliography{bibliography}
\end{document}